\pdfoutput=1
\documentclass[aps,prd,twocolumn,superscriptaddress,preprintnumbers,floatfix,nofootinbib]{revtex4-2}

\usepackage{graphicx}
\usepackage{amsmath, amssymb, mathrsfs}
\usepackage{color}
\usepackage{bm}
\usepackage[T1]{fontenc}
\usepackage[utf8]{inputenc}
\usepackage{microtype}
\usepackage[normalem]{ulem}
\usepackage[dvipsnames]{xcolor}
\usepackage[colorlinks,urlcolor=NavyBlue,citecolor=NavyBlue,linkcolor=NavyBlue,pdfusetitle]{hyperref}
\usepackage[all]{hypcap}
\usepackage{orcidlink}
\usepackage{verbatim}

\graphicspath{{figs/}}

\newcommand{\lcs}[1]{\textcolor{WildStrawberry}{#1}}
\newcommand{\Dongze}[1]{\textcolor{Green}{#1}}
\newcommand{\BB}[1]{\textcolor{blue}{#1}}

\newcommand{\tabParameters}{%
\begin{table}[t]\label{tab:parameters}
    \centering
    \begin{tabular}{c|c|c}
     Model    & NR & PN\\
     \hline
     $q$    &  $3.0001$ & $2.9975 \pm 0.0002$\\
     $M$    &  $1.00000$ & $0.99980 \pm 0.00006$
    \end{tabular}
    \caption{%
      The intrinsic parameters for the NR system and the PN fit using 22PN phase evolution. The quoted uncertainties are $1\sigma$ fit errors estimated from the numerical Jacobian of the residual at the minimum \citep[Secs.~15.5--15.6]{press2007numerical}.
    }
    \label{tab:placeholder}
\end{table}%
}

\newcommand{\figEdot}{%
\begin{figure*}[t]
\centering
\includegraphics[width=0.95\textwidth]{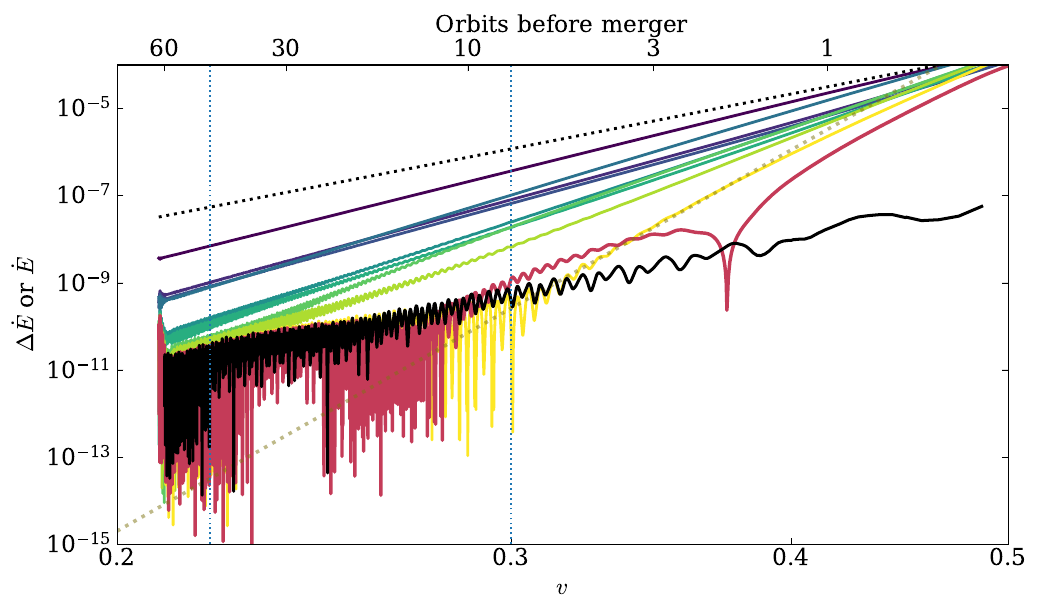}
\par
\begingroup\footnotesize\setlength{\tabcolsep}{2.5pt}
\begin{tabular}{c*{13}{c}}
\hline
& \shortstack{$1\mathrm{PN}$\\[-1pt]{\color[rgb]{0.267004,0.004874,0.329415}\rule{12pt}{1.6pt}}} & \shortstack{$1.5\mathrm{PN}$\\[-1pt]{\color[rgb]{0.278826,0.175490,0.483397}\rule{12pt}{1.6pt}}} & \shortstack{$2\mathrm{PN}$\\[-1pt]{\color[rgb]{0.229739,0.322361,0.545706}\rule{12pt}{1.6pt}}} & \shortstack{$2.5\mathrm{PN}$\\[-1pt]{\color[rgb]{0.172719,0.448791,0.557885}\rule{12pt}{1.6pt}}} & \shortstack{$3\mathrm{PN}$\\[-1pt]{\color[rgb]{0.127568,0.566949,0.550556}\rule{12pt}{1.6pt}}} & \shortstack{$3.5\mathrm{PN}$\\[-1pt]{\color[rgb]{0.157851,0.683765,0.501686}\rule{12pt}{1.6pt}}} & \shortstack{$4\mathrm{PN}$\\[-1pt]{\color[rgb]{0.369214,0.788888,0.382914}\rule{12pt}{1.6pt}}} & \shortstack{$4.5\mathrm{PN}$\\[-1pt]{\color[rgb]{0.678489,0.863742,0.189503}\rule{12pt}{1.6pt}}} & \shortstack{$6\mathrm{PN}^{*}$\\[-1pt]{\color[rgb]{0.993248,0.906157,0.143936}\rule{12pt}{1.6pt}}} & \shortstack{$22\mathrm{PN}^{*}$\\[-1pt]{\color[rgb]{0.768627,0.231373,0.345098}\rule{12pt}{1.6pt}}} & \shortstack{$6\mathrm{PN}^*$ extrap.\\[-1pt]{\color[rgb]{0.496624,0.453078,0.071968}\rule{1.2pt}{1.6pt}\hspace{1.2pt}\rule{1.2pt}{1.6pt}\hspace{1.2pt}\rule{1.2pt}{1.6pt}\hspace{1.2pt}\rule{1.2pt}{1.6pt}\hspace{1.2pt}\rule{1.2pt}{1.6pt}}} & \shortstack{NR $\dot E$\\[-1pt]{\color[rgb]{0.000000,0.000000,0.000000}\rule{1.2pt}{0.8pt}\hspace{1.2pt}\rule{1.2pt}{0.8pt}\hspace{1.2pt}\rule{1.2pt}{0.8pt}\hspace{1.2pt}\rule{1.2pt}{0.8pt}\hspace{1.2pt}\rule{1.2pt}{0.8pt}}} & \shortstack{NR error\\[-1pt]{\color[rgb]{0.000000,0.000000,0.000000}\rule{12pt}{0.8pt}}} \\
\hline
$k$ & 13 & 14 & 14 & 16 & 17 & 17 & 20 & 17 &  &  & 29 & 10 &  \\
\hline
\end{tabular}
\par\endgroup
\caption{%
Colored curves: absolute differences in energy flux $\Delta \dot{E}=|\dot{E}_{\mathrm{PN}}-\dot{E}_{\mathrm{NR}}|$ at different PN orders. %
Dotted black: the NR $\dot{E}$ from the highest resolution. %
Solid black: $\Delta \dot{E}$ between two highest NR resolutions after they are mapped to the same BMS frame, which measures the NR truncation error. %
Dotted brown line: the extrapolated 6PN residual, fitted in the region $0.32\leq v\leq0.4$. Slopes fitted between the two vertical dotted lines, $0.22\leq v\leq0.3$, and the 6PN extrapolated slope are listed in the legend, rounded to the nearest integer. The PN parameters and BMS frame are determined using 22PN phase evolution in a 5-orbit matching window starting at $v=0.21$, or around 60 orbits before merger (shortly after the junk radiation left the system).
\newline $^{*}$ \footnotesize{The fluxes above 4.5PN are incomplete at comparable masses; the additional terms are those known in the Schwarzschild EMR limit.}
}
\label{fig:Edot}
\end{figure*}%
}

\newcommand{\figvslice}{%
\begin{figure*}[t]
\centering
\includegraphics[width=0.95\textwidth]{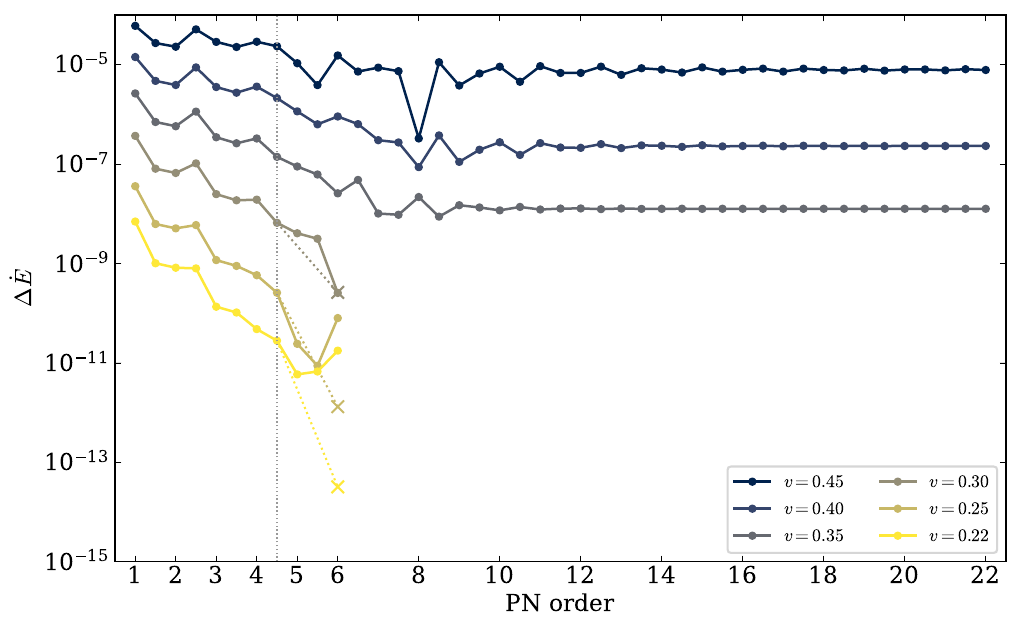}
\caption{%
  The energy flux residual $\Delta \dot{E}$ between different PN orders and NR
  at some $v$ values, with respect to the PN truncation orders.
  These are vertical slices of Fig.~\ref{fig:Edot} at different $v$. For $v\leq0.30$, points at 6.5PN and higher are omitted because the residuals are dominated by NR error. The vertical dotted line marks the highest complete PN order, 4.5PN.
  The cross markers appearing at 6PN are the extrapolated data from the dotted
  brown line of Fig.~\ref{fig:Edot}.}
\label{fig:vslice}
\end{figure*}%
}

\begin{document}

\title{Convergence of the post-Newtonian series in the quasicircular, nonprecessing, and comparable-mass regime}

\author{Dongze Sun
\orcidlink{0000-0003-0167-4392}}
\email{dzsun@caltech.edu}
\affiliation{Theoretical Astrophysics 350-17, 
California Institute of Technology, 1200 E California Boulevard, Pasadena, CA 91125, USA}
\affiliation{Max Planck Institute for Gravitational Physics (Albert Einstein Institute), Am M\"uhlenberg 1, Potsdam 14476, Germany}
\author{B\'eatrice Bonga
\orcidlink{0000-0002-5808-9517}}
\email{bbonga@science.ru.nl}
\affiliation{Institute for Mathematics, Astrophysics and Particle Physics, Radboud University, 6525 AJ Nijmegen, The Netherlands}
\affiliation{Theoretical Sciences Visiting Program, Okinawa Institute of Science and
Technology Graduate University, Onna, 904-0495, Japan}

\author{Leo C.\ Stein\,\orcidlink{0000-0001-7559-9597}}
\email[]{lcstein@olemiss.edu}
\affiliation{Department of Physics and Astronomy,
    University of Mississippi, University, MS 38677, USA}


\author{Guido Da Re
  \orcidlink{0009-0007-2006-094X}}
\affiliation{Theoretical Astrophysics 350-17, 
California Institute of Technology, 1200 E California Boulevard, Pasadena, CA 91125, USA}

\hypersetup{pdfauthor={Sun et al.}}

\date{\today}

\begin{abstract}
Post-Newtonian (PN) theory provides the analytic foundation for modeling the early inspiral of binary black holes. However, as an asymptotic series, successive PN orders do not necessarily improve agreement with the full nonlinear dynamics. 
While this has been explored in the extreme-mass-ratio (EMR) limit, comparable-mass systems most relevant to current observations have not been benchmarked as systematically at high PN order. 
We study the convergence of the PN series for non-spinning and quasi-circular systems by comparing the PN energy flux at future null infinity to a long, high-accuracy numerical relativity (NR) simulation. 
To enable a gauge-consistent comparison, we place both descriptions in the same BMS frame and calibrate the intrinsic PN parameters by fitting to the NR waveform in the early inspiral.
We find that for orbital velocities $v\lesssim0.45$, higher PN orders continue to reduce the PN--NR flux discrepancy, with EMR terms beyond 4.5PN providing further improvement in the late inspiral.
The improvement with PN order is non-monotonic with local extrema around 2.5PN and 4PN. Thus, for the comparable-mass flux comparison considered here, the first local minimum of the finite-order residual does not identify the best PN truncation order among the orders available.
As $v$ approaches $\sim 0.5$ near the innermost circular orbit, higher PN orders no longer improve the agreement between NR and PN, indicating a loss of convergence. 
These results motivate continued high-order PN calculations and clarify the NR accuracy needed to validate them.
\end{abstract}

\maketitle



\section{Introduction}
\label{sec:introduction}
Numerical relativity (NR) simulations of binary black hole mergers provide highly accurate gravitational waveforms, but their computational cost typically limits them to the late inspiral and merger-ringdown phases. To model the complete inspiral-merger-ringdown signal spanning the frequency bands of current gravitational-wave detectors, it is necessary to hybridize NR waveforms with analytic post-Newtonian (PN) waveforms, which accurately describe the early inspiral \cite{Santamaria:2010yb,MacDonald:2011ne,Boyle:2011dy,MacDonald:2012mp,Varma:2018mmi,Sadiq:2020hti}. Constructing reliable hybrid waveforms is non-trivial: it requires fixing the Bondi-Metzner-Sachs (BMS) frame of the NR waveforms to match the PN BMS frame \cite{Mitman:2022kwt,Yoo:2023spi,Sun:2024kmv} and adjusting the PN intrinsic parameters to account for the discrepancy between the PN and NR parameter definitions \cite{Sun:2024kmv,Sun:2025una}. More fundamentally, the validity of such hybrid models depends critically on (1) understanding the regime where PN theory provides a reliable approximation and where it breaks down, and (2) knowing how accurate NR needs to be, in order for hybridization to be justified.

Post-Newtonian theory is an asymptotic approximation to General Relativity in the orbital velocity $v/c$ (or equivalently in $GM/rc^2$) \cite{Futamase:1983vsr}.  Unlike convergent Taylor series, asymptotic series need not converge as more terms are added; beyond an optimal truncation order, additional terms in fact degrade rather than improve the approximation (for a pedagogical example, see \cite[Ch. 3.5]{bender1999advanced} as well as \cite[Sec. IIE]{Yunes:2008tw}).
The optimal truncation order of the PN series is velocity-dependent. At lower velocities (smaller $v/c$), the optimal truncation order is higher and one can achieve better accuracy (i.e.~approximate the full non-linear evolution as predicted by General Relativity better). Whereas at higher velocities approaching merger, the optimal truncation order is lower and the approximation becomes inherently less accurate.
This raises a crucial question for gravitational-wave modeling: have we already reached the optimal PN order for current applications, or does pushing to higher orders still yield improvements? And related, if such improvements can be made, how accurate will our NR models need to be for cross-validation?

Determining the optimal truncation order of an asymptotic series ideally requires knowledge of both the exact solution and all terms in the expansion. In the absence of either, one can still assess convergence by using an alternative high-accuracy approximation as a benchmark. For binary black holes, previous studies have investigated PN convergence in the extreme-mass-ratio (EMR) limit, where black hole perturbation theory provides highly accurate predictions that can serve as such a benchmark. To ensure meaningful comparisons between different approximation schemes, which may employ different coordinate systems and gauge choices, these studies have focused on quantities such as the gravitational-wave energy flux at null infinity, which is invariant up to boosts and BMS supertranslations.
Early work by Simone, Poisson and Will \cite{Simone:1995qu} computed the energy flux for head-on black hole collisions using both PN theory and black hole perturbation theory. Their results revealed slow PN convergence;  a feature later confirmed by high-order perturbative calculations that are  characterized by rapidly growing coefficients in successive powers of $v/c$ \cite{Fujita:2012cm,Fujita:2014eta}.
This work was followed up for quasi-circular EMR binaries in \cite{Poisson:1995vs}. Subsequent studies by Yunes and Berti \cite{Yunes:2008tw} improved upon these comparisons using a finite-order region-of-validity diagnostic, with a Kerr follow-up by Zhang, Yunes, and Berti discussing additional caveats in the error-function behavior \cite{Zhang:2011vha}, and the analysis was revisited by Sago, Fujita and Nakano in \cite{Sago:2016xsp} when higher-order PN results in the EMR limit became available. 

Here, we revisit this question in a regime where PN theory is expected to perform better: comparable mass systems \cite{Damour:2013hea}.
With increasingly accurate numerical evolutions of compact binaries now available, it is time to re-examine these convergence questions using the improved knowledge of the ``true'' numerical evolution of the system.
This expands upon early comparisons between PN and long NR waveforms initiated in~\cite{Boyle:2007ft}.

In this paper, when we refer to the ``convergence'' of the PN expansion, we do not mean convergence in the strict mathematical sense of an infinite series. Rather, we study the behavior of the finite sequence of currently known PN truncations for a specific observable, the energy flux at future null infinity, using NR as a benchmark. More precisely, at fixed $v$, we ask whether the PN–NR discrepancy decreases as additional PN terms are included. Since the PN expansion is asymptotic, such improvement need not persist to arbitrarily high order, and may eventually reverse beyond an optimal truncation order. Moreover, this behavior is quantity-dependent: different gauge-invariant observables can exhibit different large-order behavior, so the conclusions of this paper apply specifically to the energy flux.

By comparing the PN and NR energy fluxes after placing both descriptions in the same BMS frame, with the frame transformations computed from the strain and Moreschi supermomentum \cite{Moreschi:1988pc}, we find that, perhaps unsurprisingly, we have not yet saturated the utility of higher-order PN terms for gravitational-wave form modeling. Importantly, our analysis shows that the first local minimum in the finite-order flux residual does not identify the best available PN truncation order in this comparable-mass PN--NR comparison.

Nonetheless, at $v\lesssim 0.32$, or roughly 5 orbits before merger, the error in the incomplete 6PN approximation falls below the NR error. This suggests that, when constructing PN--NR hybrids with the current NR accuracy, the NR portion of the waveform need not extend deep into the inspiral.

The outline of our paper is as follows. In Sec.~\ref{sec:method}, we describe the NR and PN waveforms used in this work, the computation of the energy flux, and the procedure for placing the two descriptions in a common BMS frame while fitting the PN intrinsic parameters. In Sec.~\ref{sec:results}, we present the PN--NR energy-flux comparison, analyze the convergence behavior of the PN series at different orbital velocities, and study the dependence of our results on the choice of matching window. In Sec.~\ref{sec:disc-concl}, we discuss the implications of these results for waveform modeling and summarize possible directions for future work. Throughout, we work in geometrized units with $G=c=1$.

\section{Method}\label{sec:method}
In this section, we describe the high-accuracy NR waveform for a comparable-mass, non-precessing binary that will serve as our ``reference'' waveform. We will compute the resulting energy flux and compare this to PN results. In order to determine which PN parameters to use to compare to the NR simulation, we employ the hybridization framework introduced in Ref.~\cite{Sun:2024kmv}. This allows us to make a meaningful comparison by ensuring that both results are in the same Bondi–Metzner–Sachs (BMS) frame and that the PN parameters best match the NR result.

\subsubsection{NR waveform and energy flux}
\label{sec:NR}
We use a NR waveform computed using Cauchy-characteristic evolution (CCE) \cite{Moxon:2020gha,Moxon:2021gbv} as the reference waveform.
The Cauchy evolution is performed using the Spectral Einstein Code (SpEC) \cite{SpECwebsite} and the CCE evolution is performed
using the SpECTRE code \cite{spectrecode}.
The resulting waveform is given directly at future null infinity $\mathcal{I}^+$, along with the Bondi news, shear, and Weyl scalars.

We use the system SXS:BBH:2265 from the public SXS catalog \cite{SXSCatalog,Boyle:2019kee,Scheel:2025jct} for comparison with PN.
This system is the longest non-spinning and quasi-circular simulation in the SXS catalog to date, which has over 60 orbits available before merger.
This choice allows  comparison with the PN series at relatively small orbital velocities.
The system has total mass $M=1$ and mass ratio $q=3$, based on the NR quasi-local measurement on apparent horizons~\cite{Boyle:2019kee}.
The system is quasi-circular, but has a small residual eccentricity on the order of $10^{-5}$.

The NR energy flux at $\mathcal{I}^+$ is calculated from the strain
$h=h_{+}-ih_{\times}$ based on \cite{Ruiz:2007yx}
\begin{equation}
  \dot{E}_\infty=\oint_{\mathcal{I}^+}\frac{r^2}{16\pi}|\dot{h}|^2d\Omega
  \,.
\end{equation}
We will compute the energy flux as a function of the
orbital velocity, $\dot{E}_\infty(v)$.  The NR orbital velocity parameter is
defined as $v \equiv (GM \Omega)^{1/3}$, where the orbital frequency $\Omega$ is
measured from wave-zone data after BMS frame fixing (frame fixing is discussed
below in Sec.~\ref{sec:frame fixing}).\footnote{%
  The far-zone frequency $\Omega$ differs from the frequency that near-zone
  observers measure, denoted $\omega$.  The far-zone frequency is approximately
  $\Omega \approx \Omega_{2,2}/2$, where $\Omega_{2,2}$ is the instantaneous
  frequency of the $(2,2)$ mode -- clearly linked to asymptotic observers.  This
  leads to the two distinct PN expansion parameters,
  $x\equiv(GM\Omega/c^{3})^{2/3}$ and $y\equiv(GM\omega/c^{3})^{2/3}$, which
  differ at 4PN and higher orders.  See~\cite{Trestini:2025nzr} for more
  discussion.} %
The orbital frequency is defined as the frequency of a rotating frame in which the time-dependence
of gravitational waves is minimized~\cite{Boyle:2013nka}; or equivalently, the
frequency for a helical vector field $\partial_{t}+\Omega \partial_\phi$ that is
closest to being a Killing symmetry~\cite{Khairnar:2024rzs}.  The NR energy flux
and orbital angular velocity are calculated using the \texttt{scri}
package~\cite{scri}.

\subsubsection{PN waveform and energy flux}\label{sec:PN}
For the PN waveforms, we also consider quasi-circular, non-spinning binaries characterized by the total mass $M$ and mass ratio $q = m_1/m_2\geq1$.
The orbital dynamics and gravitational-wave phase are modeled using the TaylorT1 approximant.
In this approach, one writes the binary’s binding energy $E$, gravitational-wave energy flux to null infinity $\dot{E}_\infty$ and into the horizon $\dot{E}_\text{H}$ as PN series in the orbital velocity $v=(GM\Omega)^{1/3}$.
The binary's phase evolution is found by numerically integrating the energy-balance equation
\begin{equation}
\frac{dv}{dt}=-\frac{\dot{E}_\infty+\dot{E}_\text{H}}{M dE/dv}.
\end{equation}

Non-spinning contributions to $E$ and $\dot{E}_\infty$ for quasi-circular BBHs are complete through 4.5PN order \cite{Damour:2014jta,Bernard:2017ktp,Blanchet:2013haa,Blanchet:2023bwj}.
We include all the complete terms, supplemented by higher-order terms known in the Schwarzschild EMR limit \cite{Fujita:2012cm} up to 22PN. Orders above 4.5PN are incomplete at comparable masses and are marked by an asterisk in the figures.
We include $\dot{E}_\text{H}$ up to 4PN from Ref.~\cite{Alvi:2001mx}.

For a consistent comparison between the NR and PN energy flux, we need to know which PN parameters to use to compare to the NR result and to ensure that the BMS frame of the NR and PN setting is the same. For instance, for the PN flux, one cannot simply use the same numerical values for $M$ and $q$ as in the NR simulation, because the definition of the NR masses and PN masses have a very different origin. The NR masses are  measured quasi-locally from the apparent horizon, while the PN masses are attributes of point particles defined via asymptotic matching in body zones. To address these issues, we use the hybridization method from \cite{Sun:2024kmv} (more details below). This method requires a comparison at the level of the waveform. The result of this method is an NR waveform in the same BMS frame as the PN setting (up to numerical error) and a PN waveform with the PN parameters such that this waveform matches the NR waveform in the early phase. Once the PN parameters are determined, we can directly use the PN energy flux formulas to compute the energy flux and compare this to the NR energy flux (without the need to use the PN waveform itself to compute the PN energy flux). Thus, as a first step, we need an accurate description of the PN gravitational-wave strain modes $h_{lm}$.

For quasi-circular non-spinning binaries, the full waveform is complete through 3.5PN order, including both the oscillatory modes and the non-oscillatory memory modes \cite{Blanchet:2008je,Faye:2012we,Faye:2014fra,Favata:2008yd,Henry:2021cek,Henry:2022ccf,Cunningham:2024dog}. The dominant $(2,2)$ mode is additionally known through 4PN order \cite{Blanchet:2023sbv}. In our implementation, we include all oscillatory modes through 3.5PN order, supplemented by the 4PN correction to the $(2,2)$ mode, and include the $m=0$ memory modes through 3.5PN order. These modes are used to compute the asymptotic Poincaré charges associated with the PN waveform at $\mathcal{I}^+$. Those charges enter the frame-fixing procedure, where we align the PN and NR descriptions in a common asymptotic frame before comparing their energy fluxes.

The PN evolution including the above $E$, $\dot{E}_\infty$, $\dot{E}_\text{H}$ and strain mode terms are implemented in the code \cite{Sun:HybridizationWaveforms}, which derives from the code \texttt{GWFrames} \cite{Boyle:2013nka}. 

\subsubsection{Comparing NR and PN energy flux}
\label{sec:frame fixing}

The energy flux at $\mathcal{I}^+$ is invariant under
rotations, but not under boosts, spatial translations, or proper BMS supertranslations. These transformations change the labeling of the constant-retarded-time cross sections of $\mathcal{I}^+$, and hence modify the flux as a function of retarded time. To make meaningful comparisons between PN and NR, we must map the NR system into the PN frame.  We therefore place both descriptions in a
common BMS frame and, in practice, we choose the PN BMS frame as our reference,
since it has simple initial conditions as $u\to -\infty$, where $u$ is the
retarded time along $\mathcal{I}^{+}$.  Following Ref.~\cite{Mitman:2022kwt}, we
determine the boost that maps the NR center-of-mass charge to the PN
center-of-mass charge and apply this boost to the NR waveform.  Operationally,
the boost is found from the average time derivative of the center-of-mass
charge.  The BMS supertranslations are computed by mapping
the NR Moreschi supermomentum to the analytically-known PN Moreschi supermomentum, as detailed in
Refs.~\cite{Mitman:2022kwt,Sun:2024kmv}.

Even in the same frame, the NR and PN system have different intrinsic parameters
due to different parameter definitions and truncation errors (numerical
truncation for NR, and finite truncation in powers of $v$ in PN).
Following Ref.~\cite{Sun:2024kmv}, we therefore treat the PN intrinsic parameters as fit parameters and determine them by minimizing the fractional $L^2$ error in the waveforms. In other words, we minimize the cost function
\begin{equation}\label{eq:cost}
\mathcal{E}[h^{\mathrm{NR}}, h^{\mathrm{PN}}]=\frac{1}{2} \frac{\sum_{\ell, m} \int_{t_1}^{t_2}\left|h^{\mathrm{NR}}_{\ell m}(t) - h^{\mathrm{PN}}_{\ell m}(t;\vec{\lambda})\right|^2 d t}{\sum_{\ell, m} \int_{t_1}^{t_2}\left|h^{\mathrm{NR}}_{\ell m}(t)\right|^2 d t},
\end{equation}
where $\vec{\lambda}$ includes the intrinsic parameters $M$ and $q$, and
a time shift $\Delta t$, and a frame rotation $\Delta \phi$.
We include all $-\ell\leq m \leq \ell$ for $\ell\leq 8$ spin-weighted spherical harmonic modes in the cost function.
The cost function is evaluated in a ``matching window'' during the early inspiral.
Our default matching window is 5 orbits long and begins just after the junk radiation has left the system. We also explore alternative matching windows to probe the robustness of our conclusions, as detailed in Sec.~\ref{sec:results}.

The determination of the boost, spatial translations, supertranslations and the fit of the intrinsic parameters are performed iteratively. 
Starting from NR values as an initial guess, we alternate between updating the boost, spatial translations and supertranslations (and boosting, spatially translating and supertranslating the NR waveform into the PN frame) and updating the PN parameters $\vec{\lambda}$, until the cost function $\mathcal{E}$ converges with a fractional change smaller than $10^{-2}$. 
The final best-fit physical parameters are summarized in Tab.~\ref{tab:parameters}. 
We do not list the fitting parameters that are pure gauge (rotation, boost,
time shift, spatial translation, and 77 proper supertranslation parameters).

\tabParameters

Given the large number of (mostly gauge) fit parameters, one might worry about overfitting. In practice, however, we fit only two physical PN parameters, $(M,q)$, and effectively at a single reference value of $v$. The flux comparison is then performed over a wide range of $v$, providing a nontrivial check against overfitting; see~\cite{Sun:2024kmv} for a more systematic study.

After mapping both PN and NR to the same frame and performing the parameter optimization, we compute the NR energy flux and orbital velocity $v$ as discussed in Sec.~\ref{sec:NR}.
The NR energy flux computed this way provides our best approximation to the
``true'' energy flux of a comparable-mass, quasi-circular, non-spinning binary.
To compare PN against NR,
we then evaluate the PN energy flux at the NR orbital velocity $v$, at a variety
of PN orders.

Finally, we emphasize that both the frame-fixing and parameter-fitting procedures rely on the PN strain used here, which includes non-spinning amplitude terms through 3.5PN order and the 4PN correction to the dominant $(2,2)$ mode. 
We use phase evolution at 22PN for frame fixing and parameter fitting, and keep the resulting parameters and frame fixed when comparing the different flux truncations.
As a result, PN--NR comparisons of the energy flux at higher PN orders ($>4$) do not consistently incorporate the corresponding higher-order corrections in the strain and should be interpreted with some caution.
Nevertheless, existing evidence suggests that the difference between PN and NR is dominated by differences in the phase evolution rather than by inaccuracies in the strain amplitudes \cite{Mitman:2025tmj}, so the impact of using these strain modes on our higher-order PN comparisons should be subdominant.

\section{Results}\label{sec:results}

\figEdot

In Fig.~\ref{fig:Edot} we show the differences between the PN and NR energy flux as a function of the orbital velocity $v$ for several PN truncation orders. For reference, we have also included the total NR energy flux in black (dotted).
To quantify the small-$v$ behavior, we fit the displayed PN curves through 4.5PN in the range $0.22\leq v\leq 0.3$ (between the vertical dotted lines) using a simple linear ansatz $\log(\Delta\dot{E})=k\log(v)+b$, and extract the power-law slope $k$.
Since the PN energy flux contains an overall $v^{10}$ factor, the $N$-PN correction is expected to scale as $v^{10+2N}$, thus the PN truncation error
should be asymptotic to $\sim v^{10+2N+1}$. The fitted slopes through 4.5PN
are roughly consistent with this expectation, confirming
that we are probing the PN truncation error in this regime.
Although the $v\to 0$ asymptotics predict definite slopes, the differences between
analytical slopes and numerically fitted slopes should not be alarming given our
modest values of $0.22\leq v\leq 0.3$ and non-trivial frame-fixing and
parameter-fitting procedures.

For $v\lesssim 0.45$, increasing the PN order generally reduces the PN--NR energy-flux discrepancy, with the EMR terms beyond 4.5PN providing further improvement.
This improvement is non-monotonic and depends on the orbital velocity.
However, we expect PN to break down near the innermost circular orbit (ICO), which is known at 4PN for equal-mass non-spinning systems to be at $v\sim0.47$ \cite{Blanchet:2025agj}.
We see that as $v$ approaches $0.5$, the energy-flux differences at different PN orders cluster together, and increasing the PN order no longer improves agreement between PN and NR.
This signals the loss of convergence of the PN expansion near the ICO.

Note that the 6PN curve in Fig.~\ref{fig:Edot} exhibits large oscillations for $v\lesssim 0.32$, which obscures the clean power-law PN truncation error we expect at small $v$.
These oscillations are caused by the NR truncation error, as their magnitude aligns with the NR truncation error shown in the black curve.
To estimate the underlying PN truncation error, we therefore extrapolate the 6PN fit obtained in the range $0.32\leq v\leq0.4$ down to $0.2\leq v \leq 0.32$, shown as the brown dotted line in Fig.~\ref{fig:Edot}.
The slope of this extrapolated line is much larger than the expected $10+2N+1=23$, suggesting that the missing higher order PN terms become dominant in the region $0.32\leq v\leq0.4$.
Despite this caveat, if we take this extrapolation seriously, this indicates that at $v=0.2$, the 6PN flux is accurate at the level of $10^{-15}$, comparable to double machine precision.
In this regime, truly testing 6PN would therefore require NR simulations with much higher accuracy. Accordingly, pushing the non-spinning flux beyond 6PN is unlikely to yield measurable gains for present or upcoming detectors, while extending high-order PN results to spinning and eccentric binaries should be more beneficial in practice.

To better visualize the behavior at fixed orbital velocities, Fig.~\ref{fig:vslice} shows the energy flux difference as a function of PN order for several constant-$v$ slices.
For $v\lesssim 0.45$, the residuals generally decrease with PN order. The rate at which the residuals decrease is larger for smaller $v$, again indicating convergence.
However, the decrease is not monotonic, as there are local extrema around 2.5PN and 4PN.
The higher-order EMR terms introduce further minima. The residuals become less sensitive to the truncation order as it approaches 22PN.
This non-monotonic pattern shows that, in the present comparable-mass flux comparison, the first local minimum of the finite-order residual is not a reliable way to identify the best available truncation order.

\figvslice

We have checked the robustness of our conclusions against the choice of hybridization window.
Our default matching window is 5 orbits long and begins just after the junk radiation has left the system at $v=0.21$.
Using a shorter 3-orbit window, or shifting the 5-orbit window within the early inspiral between $0.22\leq v \leq 0.3$ leads to very similar PN--NR energy flux differences.
By contrast, a much longer 15-orbit window produces noticeably different results.
Over such a long interval the evolution of $v$ is no longer negligible. The alignment must then compromise between the (slightly) different $v(t)$ evolutions of PN and NR over a broad range of $v$, rather than enforcing agreement at an effectively fixed $v$. Consequently, the two descriptions can end the matching window at slightly different values of $v$. The inferred $\Delta\dot{E}$ then reflects not only PN truncation error but also this accumulated misalignment in $v$, which biases the flux comparison.

\section{Discussion and conclusions}
\label{sec:disc-concl}
We investigated the convergence properties of the post-Newtonian approximation
by comparing the energy flux $\dot{E}_{\infty}(v)$ from high-order PN calculations against a high-accuracy numerical relativity simulation of a non-spinning, quasi-circular binary black hole system with mass ratio $q=3$. Our analysis uses a hybridization framework that ensures both the PN and NR systems are mapped to the same BMS frame and that the PN intrinsic parameters are optimally matched to the NR evolution \cite{Sun:2024kmv}.

Our results demonstrate that, for orbital velocities $v \lesssim 0.45$, increasing the PN order generally reduces the PN--NR energy-flux discrepancy. This improvement continues when EMR terms beyond 4.5PN are included, although the dependence on PN order remains non-monotonic through 22PN.
The energy flux differences scale approximately as the expected power law $\Delta\dot{E} \sim v^{10+2N+1}$ when truncating at $N$-PN order. The improvement with PN order breaks down as $v$ approaches $\sim 0.5$ near the ICO, where all PN orders cluster together and higher-order terms no longer improve agreement with NR.

Our analysis shows that, in this comparable-mass flux comparison, the first local minimum in the finite-order residual does not identify the best available PN truncation order. This should be interpreted as a limitation of directly transferring the EMR finite-order region-of-validity behavior to the present setting, not as a contradiction of the EMR analysis of Refs.~\cite{Yunes:2008tw,Zhang:2011vha}. We observe non-monotonic improvement between PN and NR with local extrema around 2.5PN and 4PN, indicating that the asymptotic nature of the PN expansion is more subtle than a simple monotonic approach to optimal truncation. 

We further show that at $v\lesssim 0.32$, or roughly 5 orbits before merger, the error in the incomplete 6PN approximation falls below the NR error. This suggests that, when constructing PN--NR hybrids for accurate waveform modeling, the NR portion of the waveform need not extend deep into the inspiral of quasi-circular black hole binaries. This of course assumes that, at earlier times, the missing terms at 5, 5.5, and 6PN do not reverse the general trends we observed.\footnote{A complete calculation of the (point-mass) 5PN contributions would of course be particularly valuable for tidal deformability inference, since the leading adiabatic tidal effects enter the phasing at 5PN order and are therefore degenerate with any unknown point-mass terms at the same order. If the point-mass 5PN terms are not fully known, any mismatch at 5PN can bias the tidal signal, directly limiting how cleanly Love numbers can be extracted from waveform data.}

A striking result emerges when we extrapolate the incomplete 6PN behavior to lower velocities: at $v \approx 0.2$, the 6PN approximation seems to achieve accuracy at the level of $\sim 10^{-15}$ in the energy flux.
Entertaining this extrapolation, this suggests that to meaningfully test 6PN theory or to justify hybridization with 6PN waveforms at such low velocities would therefore require NR simulations with absolute energy-flux errors approaching $10^{-15}$---comparable to double machine precision. This represents a formidable challenge that pushes well beyond the capabilities of current NR codes. In this sense, pushing the non-spinning flux beyond 6PN is unlikely to produce observable gains for present or near-future detectors, whereas extending high-order PN information to spinning and eccentric systems may yield greater practical benefits.

These results have practical implications for gravitational-wave astronomy. Current waveform models for LIGO-Virgo-KAGRA observations typically employ PN approximations at 3.5PN or 4PN order. Our findings indicate that pushing to higher PN orders -- 5PN and incomplete 6PN -- could yield measurable improvements in waveform accuracy, especially for signals that span the early inspiral regime where $v\lesssim 0.4$. This will be all the more important for future observations in which we expect to see longer duration signals. However, realizing these improvements requires not only the analytical calculation of higher-order PN terms but also sufficiently accurate NR simulations to properly calibrate and validate the resulting hybrid waveforms.

\subsection{Future directions}
\label{sec:future-work}
Several extensions of this work would be valuable. First, our current analysis relies on an optimization-based approach to match PN and NR parameters, minimizing the waveform mismatch over a chosen hybridization window. While this approach is effective, it carries the risk of overfitting the PN parameters to compensate for modeling errors rather than capturing genuine physical agreement. We plan to develop alternative matching procedures based on first-principle parameter mappings—for instance, relating the PN point-particle masses to NR quasi-local horizon masses through well-defined theoretical prescriptions \cite{Sun:2025una}. Such an approach would provide a more robust foundation for PN--NR comparisons and reduce sensitivity to the choice of matching window.

Second, our present study focuses on non-spinning, quasi-circular
binaries. Extending this analysis to more general systems—including spinning and
eccentric configurations—represents an important next step. For spinning
systems, understanding PN convergence becomes more complex due to the richer
parameter space and the interplay between orbital and spin dynamics. Recent
advances in high-order spin-dependent PN calculations \cite{Cho:2022syn,Khalil:2021fpm,Kim:2021rfj} and improvements in NR accuracy for spinning systems \cite{Scheel:2025jct,Ferguson:2023vta} now make such comparisons feasible. Similarly,
eccentric binary simulations have matured considerably, and eccentric PN theory
has been developed to high orders in some limits (although more developments
here are certainly welcome). Systematic PN--NR comparisons for eccentric systems
would both benchmark the eccentric PN approximation and probe if the
convergence properties we observe for circular orbits persist when eccentricity
is present.

However, extending these comparisons to eccentric systems is substantially more challenging. In this work, we took advantage of the fact that at fixed $(M,q)$, the time evolution of a quasicircular non-spinning BBH naturally runs through a range of values of $v$, which we can take as the asymptotic expansion parameter.  
By contrast, in
eccentric systems, the instantaneous relative orbital velocity $v$ is not a suitable
expansion parameter---binaries with different semimajor axis and eccentricity $(a,e)$ can attain the same $v$ at different orbital phases between pericenter and
 apocenter. A more appropriate choice is the mean motion $n$, leading to the
expansion parameter $x\equiv(GM K n/c^{3})^{2/3}$ which reduces to
$x\to v^{2}/c^{2}$ in the quasicircular limit (here $2\pi K$ gives the advance
of pericenter per radial orbit).  An individual eccentric inspiral then traces only a single path through $(x,e)$ space. By contrast, a quasicircular inspiral sweeps across the entire one-dimensional range of $v$. Demonstrating convergence with eccentricity will therefore require a family of binaries covering initial eccentricities $0\le e<1$.

These extensions would serve the dual purpose of benchmarking PN theory against NR in broader regions of parameter space while simultaneously using high-order PN results to validate and test the accuracy of state-of-the-art NR simulations. As NR and PN methods continue to improve in tandem, such cross-validation becomes increasingly valuable for ensuring the reliability of gravitational-wave models across the full parameter space relevant for current and future detectors.


\acknowledgments
%
%
We would like to thank the organizers of the workshop  ``Mathematical Methods for the General Relativistic Two-body Problem'' at the Institute for Mathematical Sciences (IMS) at the National University of Singapore (NUS), which provided the environment that stimulated this idea.
We would also like to thank D. Trestini for valuable feedback on an earlier version of this manuscript.
The work at Caltech was supported in part by the Sherman Fairchild Foundation, and
by NSF Grants PHY-2309211, PHY-2309231, and OAC-2513339, and
NASA award 80NSSC26K0340.
Computations for this work were preformed on the Resnick High Performance Computing (HPC) Cluster at the Caltech High Performance Computing Center.
This research was conducted while BB was visiting the Okinawa Institute of Science and
Technology (OIST) through the Theoretical Sciences Visiting Program (TSVP).
L.C.S. was supported in part by NSF CAREER award PHY--2047382 and a Sloan Foundation Research Fellowship.


\bibliographystyle{JHEP}
\bibliography{PNconvergence}

@article{Damour:2013hea,
    author = "Damour, Thibault",
    title = "{The general relativistic two body problem}",
    eprint = "1312.3505",
    archivePrefix = "arXiv",
    primaryClass = "gr-qc",
    month = "12",
    year = "2013"
}

@book{bender1999advanced,
  title={Advanced Mathematical Methods for Scientists and Engineers: Asymptotic Methods and Perturbation Theory},
  author={Bender, C.M. and Orszag, S.A.},
  isbn={9780387989310},
  lccn={99044783},
  series={Advanced Mathematical Methods for Scientists and Engineers},
  urlIGNORE={https://books.google.com/books?id=-yQXwhE6iWMC},
  year={1999},
  publisher={Springer}
}

@article{Blanchet:2023bwj,
    author = "Blanchet, Luc and Faye, Guillaume and Henry, Quentin and Larrouturou, Fran{\c{c}}ois and Trestini, David",
    title = "{Gravitational-Wave Phasing of Quasicircular Compact Binary Systems to the Fourth-and-a-Half Post-Newtonian Order}",
    eprint = "2304.11185",
    archivePrefix = "arXiv",
    primaryClass = "gr-qc",
    reportNumber = "DESY-23-043",
    doi = "10.1103/PhysRevLett.131.121402",
    journal = "Phys. Rev. Lett.",
    volume = "131",
    number = "12",
    pages = "121402",
    year = "2023"
}

@article{Futamase:1983vsr,
    author = "Futamase, T. and Schutz, Bernard F.",
    title = "{Newtonian and post-Newtonian approximations are asymptotic to general relativity}",
    doi = "10.1103/PhysRevD.28.2363",
    journal = "Phys. Rev. D",
    volume = "28",
    number = "10",
    pages = "2363",
    year = "1983"
}

@article{Yunes:2008tw,
    author = "Yunes, Nicolas and Berti, Emanuele",
    title = "{Accuracy of the post-Newtonian approximation: Optimal asymptotic expansion for quasicircular, extreme-mass ratio inspirals}",
    eprint = "0803.1853",
    archivePrefix = "arXiv",
    primaryClass = "gr-qc",
    reportNumber = "IGC-08-3-1",
    doi = "10.1103/PhysRevD.77.124006",
    journal = "Phys. Rev. D",
    volume = "77",
    pages = "124006",
    year = "2008",
    note = "[Erratum: Phys.Rev.D 83, 109901 (2011)]"
}

@article{Sago:2016xsp,
    author = "Sago, Norichika and Fujita, Ryuichi and Nakano, Hiroyuki",
    title = "{Accuracy of the Post-Newtonian Approximation for Extreme-Mass Ratio Inspirals from Black-hole Perturbation Approach}",
    eprint = "1601.02174",
    archivePrefix = "arXiv",
    primaryClass = "gr-qc",
    doi = "10.1103/PhysRevD.93.104023",
    journal = "Phys. Rev. D",
    volume = "93",
    number = "10",
    pages = "104023",
    year = "2016"
}

@article{Poisson:1995vs,
    author = "Poisson, Eric",
    title = "{Gravitational radiation from a particle in circular orbit around a black hole. 6. Accuracy of the postNewtonian expansion}",
    eprint = "gr-qc/9505030",
    archivePrefix = "arXiv",
    doi = "10.1103/PhysRevD.52.5719",
    journal = "Phys. Rev. D",
    volume = "52",
    pages = "5719--5723",
    year = "1995",
    note = "[Addendum: Phys.Rev.D 55, 7980--7981 (1997)]"
}

@article{Simone:1995qu,
    author = "Simone, Liliana E. and Poisson, Eric and Will, Clifford M.",
    title = "{Headon collision of compact objects in general relativity: Comparison of postNewtonian and perturbation approaches}",
    eprint = "gr-qc/9506080",
    archivePrefix = "arXiv",
    doi = "10.1103/PhysRevD.52.4481",
    journal = "Phys. Rev. D",
    volume = "52",
    pages = "4481--4496",
    year = "1995"
}

@article{Fujita:2014eta,
    author = "Fujita, Ryuichi",
    title = "{Gravitational Waves from a Particle in Circular Orbits around a Rotating Black Hole to the 11th Post-Newtonian Order}",
    eprint = "1412.5689",
    archivePrefix = "arXiv",
    primaryClass = "gr-qc",
    doi = "10.1093/ptep/ptv012",
    journal = "PTEP",
    volume = "2015",
    number = "3",
    pages = "033E01",
    year = "2015"
}

@software{spectrecode,
    author = "Deppe, Nils and Throwe, William and Kidder, Lawrence E. and Vu,
Nils L. and H\'ebert, Fran\c{c}ois and Moxon, Jordan and Armaza, Crist\'obal and
Bonilla, Marceline S. and Kim, Yoonsoo and Kumar, Prayush and Lovelace, Geoffrey
and Macedo, Alexandra and Nelli, Kyle C. and O'Shea, Eamonn and Pfeiffer, Harald
P. and Scheel, Mark A. and Teukolsky, Saul A. and Wittek, Nikolas A. and 
others",
    title = "\texttt{SpECTRE v2023.05.16}",
    version = "2023.05.16",
    publisher = "Zenodo",
    doi = "10.5281/zenodo.7942177",
    howpublished = "\url{https://spectre-code.org}",
    license = "MIT",
    year = "2023",
    month = "5"
}

@misc{SpECwebsite,
  title =   {The {S}pectral {E}instein {C}ode},
  howpublished = "\url{http://www.black-holes.org/SpEC.html}"
}

@misc{SXSCatalog,
  author = {{SXS Collaboration}},
  title = {The {SXS} Collaboration catalog of gravitational waveforms},
  howpublished = "\url{http://www.black-holes.org/waveforms}",
}

@software{scri,
    author = {Boyle, Michael and Iozzo, Dante and Stein, Leo and Khairnar, Aniket and Rüter, Hannes and Scheel, Mark and Varma, Vijay and Mitman, Keefe},
    doi = {10.5281/zenodo.4041971},
    month = nov,
    title = {{scri}},
    howpublished = {\url{https://github.com/moble/scri}},
    version = {2022.8.8},
    year = {2023}
}

@misc{Sun:HybridizationWaveforms,
  author = "Dongze Sun",
  title = {{NRPNHybridization}},
  howpublished = "\url{https://github.com/dongzesun/NRPNHybridization}",
}

@article{Santamaria:2010yb,
    author = "Santamaria, L. and others",
    title = "{Matching post-Newtonian and numerical relativity waveforms: systematic errors and a new phenomenological model for non-precessing black hole binaries}",
    eprint = "1005.3306",
    archivePrefix = "arXiv",
    primaryClass = "gr-qc",
    reportNumber = "LIGO-P1000048, AEI-2010-122",
    doi = "10.1103/PhysRevD.82.064016",
    journal = "Phys. Rev. D",
    volume = "82",
    pages = "064016",
    year = "2010"
}

@article{MacDonald:2011ne,
      author         = "MacDonald, Ilana and Nissanke, Samaya and Pfeiffer,
                        Harald P. and Pfeiffer, Harald P.",
      title          = "{Suitability of post-Newtonian/numerical-relativity
                        hybrid waveforms for gravitational wave detectors}",
      booktitle      = "{Theory meets data analysis at comparable and extreme
                        mass ratios. Proceedings, Conference, NRDA/CAPRA 2010,
                        Waterloo, Canada, June 20-26, 2010}",
      journal        = "Class. Quant. Grav.",
      volume         = "28",
      year           = "2011",
      pages          = "134002",
      doi            = "10.1088/0264-9381/28/13/134002",
      eprint         = "1102.5128",
      archivePrefix  = "arXiv",
      primaryClass   = "gr-qc",
      SLACcitation   = "%%CITATION = ARXIV:1102.5128;%%"
}

@article{Boyle:2011dy,
      author         = "Boyle, Michael",
      title          = "{Uncertainty in hybrid gravitational waveforms:
                        Optimizing initial orbital frequencies for binary
                        black-hole simulations}",
      journal        = "Phys. Rev.",
      volume         = "D84",
      year           = "2011",
      pages          = "064013",
      doi            = "10.1103/PhysRevD.84.064013",
      eprint         = "1103.5088",
      archivePrefix  = "arXiv",
      primaryClass   = "gr-qc",
      SLACcitation   = "%%CITATION = ARXIV:1103.5088;%%"
}

@article{MacDonald:2012mp,
      author         = "MacDonald, Ilana and Mroue, Abdul H. and Pfeiffer, Harald
                        P. and Boyle, Michael and Kidder, Lawrence E. and Scheel,
                        Mark A. and Szilagyi, Bela and Taylor, Nicholas W.",
      title          = "{Suitability of hybrid gravitational waveforms for
                        unequal-mass binaries}",
      journal        = "Phys. Rev.",
      volume         = "D87",
      year           = "2013",
      number         = "2",
      pages          = "024009",
      doi            = "10.1103/PhysRevD.87.024009",
      eprint         = "1210.3007",
      archivePrefix  = "arXiv",
      primaryClass   = "gr-qc",
      SLACcitation   = "%%CITATION = ARXIV:1210.3007;%%"
}

@article{Varma:2018mmi,
      author         = "Varma, Vijay and Field, Scott E. and Scheel, Mark A. and
                        Blackman, Jonathan and Kidder, Lawrence E. and Pfeiffer,
                        Harald P.",
      title          = "{Surrogate model of hybridized numerical relativity
                        binary black hole waveforms}",
      journal        = "Phys. Rev.",
      volume         = "D99",
      year           = "2019",
      number         = "6",
      pages          = "064045",
      doi            = "10.1103/PhysRevD.99.064045",
      eprint         = "1812.07865",
      archivePrefix  = "arXiv",
      primaryClass   = "gr-qc",
      SLACcitation   = "%%CITATION = ARXIV:1812.07865;%%"
}

@article{Sadiq:2020hti,
    author = "Sadiq, Jam and Zlochower, Yosef and O'Shaughnessy, Richard and Lange, Jacob",
    title = "{Hybrid waveforms for generic precessing binaries for gravitational-wave data analysis}",
    eprint = "2001.07109",
    archivePrefix = "arXiv",
    primaryClass = "gr-qc",
    doi = "10.1103/PhysRevD.102.024012",
    journal = "Phys. Rev. D",
    volume = "102",
    number = "2",
    pages = "024012",
    year = "2020"
}

@article{Moreschi:1988pc,
    author = "Moreschi, O. M.",
    title = "{Supercenter of Mass System at Future Null Infinity}",
    doi = "10.1088/0264-9381/5/3/004",
    journal = "Class. Quant. Grav.",
    volume = "5",
    pages = "423--435",
    year = "1988"
}

@article{Trestini:2025nzr,
    author = "Trestini, David",
    title = "{Schott term in the binding energy for compact binaries on circular orbits at fourth post-Newtonian order}",
    eprint = "2504.13245",
    archivePrefix = "arXiv",
    primaryClass = "gr-qc",
    doi = "10.1103/lsbb-sv45",
    journal = "Phys. Rev. D",
    volume = "112",
    number = "2",
    pages = "024076",
    year = "2025"
}

@article{Khairnar:2024rzs,
    author = "Khairnar, Aniket and Stein, Leo C. and Boyle, Michael",
    title = "{Approximate helical symmetry in compact binaries}",
    eprint = "2410.16373",
    archivePrefix = "arXiv",
    primaryClass = "gr-qc",
    doi = "10.1103/PhysRevD.111.024072",
    journal = "Phys. Rev. D",
    volume = "111",
    number = "2",
    pages = "024072",
    year = "2025"
}

@article{Moxon:2021gbv,
    author = "Moxon, Jordan and Scheel, Mark A. and Teukolsky, Saul A. and Deppe, Nils and Fischer, Nils and H{\'e}bert, Francois and Kidder, Lawrence E. and Throwe, William",
    title = "{SpECTRE Cauchy-characteristic evolution system for rapid, precise waveform extraction}",
    eprint = "2110.08635",
    archivePrefix = "arXiv",
    primaryClass = "gr-qc",
    doi = "10.1103/PhysRevD.107.064013",
    journal = "Phys. Rev. D",
    volume = "107",
    number = "6",
    pages = "064013",
    year = "2023"
}

@article{Bernard:2017ktp,
    author = "Bernard, Laura and Blanchet, Luc and Faye, Guillaume and Marchand, Tanguy",
    title = "{Center-of-Mass Equations of Motion and Conserved Integrals of Compact Binary Systems at the Fourth Post-Newtonian Order}",
    eprint = "1711.00283",
    archivePrefix = "arXiv",
    primaryClass = "gr-qc",
    doi = "10.1103/PhysRevD.97.044037",
    journal = "Phys. Rev. D",
    volume = "97",
    number = "4",
    pages = "044037",
    year = "2018"
}

@article{Scheel:2025jct,
    author = "Scheel, Mark A. and others",
    title = "{The SXS collaboration{\textquoteright}s third catalog of binary black hole simulations}",
    eprint = "2505.13378",
    archivePrefix = "arXiv",
    primaryClass = "gr-qc",
    doi = "10.1088/1361-6382/adfd34",
    journal = "Class. Quant. Grav.",
    volume = "42",
    number = "19",
    pages = "195017",
    year = "2025"
}

@article{Mitman:2022kwt,
    author = "Mitman, Keefe and others",
    title = "{Fixing the BMS frame of numerical relativity waveforms with BMS charges}",
    eprint = "2208.04356",
    archivePrefix = "arXiv",
    primaryClass = "gr-qc",
    doi = "10.1103/PhysRevD.106.084029",
    journal = "Phys. Rev. D",
    volume = "106",
    number = "8",
    pages = "084029",
    year = "2022"
}

@article{Yoo:2023spi,
    author = "Yoo, Jooheon and others",
    title = "{Numerical relativity surrogate model with memory effects and post-Newtonian hybridization}",
    eprint = "2306.03148",
    archivePrefix = "arXiv",
    primaryClass = "gr-qc",
    doi = "10.1103/PhysRevD.108.064027",
    journal = "Phys. Rev. D",
    volume = "108",
    number = "6",
    pages = "064027",
    year = "2023"
}

@article{Boyle:2019kee,
    author = "Boyle, Michael and others",
    title = "{The SXS Collaboration catalog of binary black hole simulations}",
    eprint = "1904.04831",
    archivePrefix = "arXiv",
    primaryClass = "gr-qc",
    doi = "10.1088/1361-6382/ab34e2",
    journal = "Class. Quant. Grav.",
    volume = "36",
    number = "19",
    pages = "195006",
    year = "2019"
}

@article{Moxon:2020gha,
    author = "Moxon, Jordan and Scheel, Mark A. and Teukolsky, Saul A.",
    title = "{Improved Cauchy-characteristic evolution system for high-precision numerical relativity waveforms}",
    eprint = "2007.01339",
    archivePrefix = "arXiv",
    primaryClass = "gr-qc",
    doi = "10.1103/PhysRevD.102.044052",
    journal = "Phys. Rev. D",
    volume = "102",
    number = "4",
    pages = "044052",
    year = "2020"
}

@article{Blanchet:2013haa,
    author = "Blanchet, Luc",
    title = "{Post-Newtonian Theory for Gravitational Waves}",
    eprint = "1310.1528",
    archivePrefix = "arXiv",
    primaryClass = "gr-qc",
    doi = "10.12942/lrr-2014-2",
    journal = "Living Rev. Rel.",
    volume = "17",
    pages = "2",
    year = "2014"
}

@article{Faye:2012we,
    author = "Faye, Guillaume and Marsat, Sylvain and Blanchet, Luc and Iyer, Bala R.",
    title = "{The third and a half post-Newtonian gravitational wave quadrupole mode for quasi-circular inspiralling compact binaries}",
    eprint = "1204.1043",
    archivePrefix = "arXiv",
    primaryClass = "gr-qc",
    doi = "10.1088/0264-9381/29/17/175004",
    journal = "Class. Quant. Grav.",
    volume = "29",
    pages = "175004",
    year = "2012"
}

@article{Sun:2024kmv,
    author = "Sun, Dongze and Boyle, Michael and Mitman, Keefe and Scheel, Mark A. and Stein, Leo C. and Teukolsky, Saul A. and Varma, Vijay",
    title = "{Optimizing post-Newtonian parameters and fixing the BMS frame for numerical-relativity waveform hybridizations}",
    eprint = "2403.10278",
    archivePrefix = "arXiv",
    primaryClass = "gr-qc",
    doi = "10.1103/PhysRevD.110.104076",
    journal = "Phys. Rev. D",
    volume = "110",
    number = "10",
    pages = "104076",
    year = "2024"
}

@article{Ruiz:2007yx,
    author = "Ruiz, Milton and Takahashi, Ryoji and Alcubierre, Miguel and Nunez, Dario",
    title = "{Multipole expansions for energy and momenta carried by gravitational waves}",
    eprint = "0707.4654",
    archivePrefix = "arXiv",
    primaryClass = "gr-qc",
    doi = "10.1007/s10714-007-0570-8",
    journal = "Gen. Rel. Grav.",
    volume = "40",
    pages = "2467",
    year = "2008"
}

@article{Faye:2014fra,
    author = "Faye, Guillaume and Blanchet, Luc and Iyer, Bala R.",
    title = "{Non-linear multipole interactions and gravitational-wave octupole modes for inspiralling compact binaries to third-and-a-half post-Newtonian order}",
    eprint = "1409.3546",
    archivePrefix = "arXiv",
    primaryClass = "gr-qc",
    doi = "10.1088/0264-9381/32/4/045016",
    journal = "Class. Quant. Grav.",
    volume = "32",
    number = "4",
    pages = "045016",
    year = "2015"
}

@article{Blanchet:2023sbv,
    author = "Blanchet, Luc and Faye, Guillaume and Henry, Quentin and Larrouturou, Fran{\c{c}}ois and Trestini, David",
    title = "{Gravitational-wave flux and quadrupole modes from quasicircular nonspinning compact binaries to the fourth post-Newtonian order}",
    eprint = "2304.11186",
    archivePrefix = "arXiv",
    primaryClass = "gr-qc",
    reportNumber = "DESY-23-044",
    doi = "10.1103/PhysRevD.108.064041",
    journal = "Phys. Rev. D",
    volume = "108",
    number = "6",
    pages = "064041",
    year = "2023"
}

@article{Favata:2008yd,
    author = "Favata, Marc",
    title = "{Post-Newtonian corrections to the gravitational-wave memory for quasi-circular, inspiralling compact binaries}",
    eprint = "0812.0069",
    archivePrefix = "arXiv",
    primaryClass = "gr-qc",
    doi = "10.1103/PhysRevD.80.024002",
    journal = "Phys. Rev. D",
    volume = "80",
    pages = "024002",
    year = "2009"
}

@article{Mitman:2025tmj,
    author = "Mitman, Keefe and Stein, Leo C. and Boyle, Michael and Deppe, Nils and Kidder, Lawrence E. and Pfeiffer, Harald P. and Scheel, Mark A.",
    title = "{Length dependence of waveform mismatch: a caveat on waveform accuracy}",
    eprint = "2502.14025",
    archivePrefix = "arXiv",
    primaryClass = "gr-qc",
    doi = "10.1088/1361-6382/add8d9",
    journal = "Class. Quant. Grav.",
    volume = "42",
    number = "11",
    pages = "117001",
    year = "2025"
}

@article{Blanchet:2008je,
    author = "Blanchet, Luc and Faye, Guillaume and Iyer, Bala R. and Sinha, Siddhartha",
    title = "{The Third post-Newtonian gravitational wave polarisations and associated spherical harmonic modes for inspiralling compact binaries in quasi-circular orbits}",
    eprint = "0802.1249",
    archivePrefix = "arXiv",
    primaryClass = "gr-qc",
    doi = "10.1088/0264-9381/25/16/165003",
    journal = "Class. Quant. Grav.",
    volume = "25",
    pages = "165003",
    year = "2008",
    note = "[Erratum: Class.Quant.Grav. 29, 239501 (2012)]"
}

@article{Alvi:2001mx,
    author = "Alvi, Kashif",
    title = "{Energy and angular momentum flow into a black hole in a binary}",
    eprint = "gr-qc/0107080",
    archivePrefix = "arXiv",
    doi = "10.1103/PhysRevD.64.104020",
    journal = "Phys. Rev. D",
    volume = "64",
    pages = "104020",
    year = "2001"
}

@article{Damour:2014jta,
    author = {Damour, Thibault and Jaranowski, Piotr and Sch{\"a}fer, Gerhard},
    title = "{Nonlocal-in-time action for the fourth post-Newtonian conservative dynamics of two-body systems}",
    eprint = "1401.4548",
    archivePrefix = "arXiv",
    primaryClass = "gr-qc",
    doi = "10.1103/PhysRevD.89.064058",
    journal = "Phys. Rev. D",
    volume = "89",
    number = "6",
    pages = "064058",
    year = "2014"
}

@article{Boyle:2013nka,
    author = "Boyle, Michael",
    title = "{Angular velocity of gravitational radiation from precessing binaries and the corotating frame}",
    eprint = "1302.2919",
    archivePrefix = "arXiv",
    primaryClass = "gr-qc",
    doi = "10.1103/PhysRevD.87.104006",
    journal = "Phys. Rev. D",
    volume = "87",
    number = "10",
    pages = "104006",
    year = "2013"
}

@article{Fujita:2012cm,
    author = "Fujita, Ryuichi",
    title = "{Gravitational Waves from a Particle in Circular Orbits around a Schwarzschild Black Hole to the 22nd Post-Newtonian Order}",
    eprint = "1211.5535",
    archivePrefix = "arXiv",
    primaryClass = "gr-qc",
    doi = "10.1143/PTP.128.971",
    journal = "Prog. Theor. Phys.",
    volume = "128",
    pages = "971--992",
    year = "2012"
}

@book{press2007numerical,
  title     = {Numerical Recipes: The Art of Scientific Computing},
  author    = {Press, William H. and Teukolsky, Saul A. and Vetterling, William T. and Flannery, Brian P.},
  edition   = {3},
  publisher = {Cambridge University Press},
  year      = {2007}
}

@article{Boyle:2007ft,
    author = "Boyle, Michael and Brown, Duncan A. and Kidder, Lawrence E. and Mroue, Abdul H. and Pfeiffer, Harald P. and Scheel, Mark A. and Cook, Gregory B. and Teukolsky, Saul A.",
    title = "{High-accuracy comparison of numerical relativity simulations with post-Newtonian expansions}",
    eprint = "0710.0158",
    archivePrefix = "arXiv",
    primaryClass = "gr-qc",
    doi = "10.1103/PhysRevD.76.124038",
    journal = "Phys. Rev. D",
    volume = "76",
    pages = "124038",
    year = "2007"
}

@article{Blanchet:2025agj,
    author = "Blanchet, Luc and Langlois, David and Ligout, Etienne",
    title = "{Innermost stable circular orbit of arbitrary-mass compact binaries at fourth post-Newtonian order}",
    eprint = "2505.01278",
    archivePrefix = "arXiv",
    primaryClass = "gr-qc",
    doi = "10.1103/mtv7-lkv8",
    journal = "Phys. Rev. D",
    volume = "112",
    number = "6",
    pages = "064025",
    year = "2025"
}

@article{Cunningham:2024dog,
    author = "Cunningham, Kevin and Kavanagh, Chris and Pound, Adam and Trestini, David and Warburton, Niels and Neef, Jakob",
    title = "{Gravitational memory: new results from post-Newtonian and self-force theory}",
    eprint = "2410.23950",
    archivePrefix = "arXiv",
    primaryClass = "gr-qc",
    doi = "10.1088/1361-6382/adbc3d",
    journal = "Class. Quant. Grav.",
    volume = "42",
    number = "13",
    pages = "135009",
    year = "2025",
    note = "[Addendum: Class.Quant.Grav. 42, 199401 (2025)]"
}

@article{Ferguson:2023vta,
    author = "Ferguson, Deborah and others",
    title = "{Second MAYA catalog of binary black hole numerical relativity waveforms}",
    eprint = "2309.00262",
    archivePrefix = "arXiv",
    primaryClass = "gr-qc",
    reportNumber = "UTWI-32-2023",
    doi = "10.1103/gk7x-9cds",
    journal = "Phys. Rev. D",
    volume = "112",
    number = "4",
    pages = "044043",
    year = "2025"
}

@article{Sun:2025una,
    author = "Sun, Dongze and Stein, Leo C.",
    title = "{Parameter matching between horizon quasi-local and point-particle definitions at 1PN for quasi-circular and non spinning BBH systems in harmonic gauge}",
    eprint = "2510.25618",
    archivePrefix = "arXiv",
    primaryClass = "gr-qc",
    month = "10",
    year = "2025"
}

@article{Kim:2021rfj,
    author = "Kim, Jung-Wook and Levi, Mich{\`e}le and Yin, Zhewei",
    title = "{Quadratic-in-spin interactions at fifth post-Newtonian order probe new physics}",
    eprint = "2112.01509",
    archivePrefix = "arXiv",
    primaryClass = "hep-th",
    doi = "10.1016/j.physletb.2022.137410",
    journal = "Phys. Lett. B",
    volume = "834",
    pages = "137410",
    year = "2022"
}

@article{Cho:2022syn,
    author = "Cho, Gihyuk and Porto, Rafael A. and Yang, Zixin",
    title = "{Gravitational radiation from inspiralling compact objects: Spin effects to the fourth post-Newtonian order}",
    eprint = "2201.05138",
    archivePrefix = "arXiv",
    primaryClass = "gr-qc",
    reportNumber = "DESY-22-004, ET-0001A-22, DESY-22-004; ET-0001A-22",
    doi = "10.1103/PhysRevD.106.L101501",
    journal = "Phys. Rev. D",
    volume = "106",
    number = "10",
    pages = "L101501",
    year = "2022"
}

@article{Khalil:2021fpm,
    author = "Khalil, Mohammed",
    title = "{Gravitational spin-orbit dynamics at the fifth-and-a-half post-Newtonian order}",
    eprint = "2110.12813",
    archivePrefix = "arXiv",
    primaryClass = "gr-qc",
    doi = "10.1103/PhysRevD.104.124015",
    journal = "Phys. Rev. D",
    volume = "104",
    number = "12",
    pages = "124015",
    year = "2021"
}

@article{Henry:2022ccf,
    author = "Henry, Quentin",
    title = "{Complete gravitational-waveform amplitude modes for quasicircular compact binaries to the 3.5PN order}",
    eprint = "2210.15602",
    archivePrefix = "arXiv",
    primaryClass = "gr-qc",
    doi = "10.1103/PhysRevD.107.044057",
    journal = "Phys. Rev. D",
    volume = "107",
    number = "4",
    pages = "044057",
    year = "2023"
}

@article{Henry:2021cek,
    author = "Henry, Quentin and Faye, Guillaume and Blanchet, Luc",
    title = "{The current-type quadrupole moment and gravitational-wave mode ({\ensuremath{\ell}}, m) = (2, 1) of compact binary systems at the third post-Newtonian order}",
    eprint = "2105.10876",
    archivePrefix = "arXiv",
    primaryClass = "gr-qc",
    doi = "10.1088/1361-6382/ac1850",
    journal = "Class. Quant. Grav.",
    volume = "38",
    number = "18",
    pages = "185004",
    year = "2021"
}

@article{Zhang:2011vha,
    author = "Zhang, Zhongyang and Yunes, Nicolas and Berti, Emanuele",
    title = "{Accuracy of the post-Newtonian approximation. II. Optimal asymptotic expansion of the energy flux for quasicircular, extreme mass-ratio inspirals into a Kerr black hole}",
    eprint = "1103.6041",
    archivePrefix = "arXiv",
    primaryClass = "gr-qc",
    doi = "10.1103/PhysRevD.84.024029",
    journal = "Phys. Rev. D",
    volume = "84",
    pages = "024029",
    year = "2011"
}

\end{document}